\documentstyle[prl,aps,floats]{revtex}
\input{psfig.tex}

\tighten
\begin{document} 
\draft

\newcommand{\be}{\begin{equation}} \newcommand{\ee}{\end{equation}}
\newcommand{\eqn}{\label}\newcommand{\bel}{\begin{equation}\label}
\def\thf{\baselineskip=\normalbaselineskip\multiply\baselineskip
by 7\divide\baselineskip by 6}
\def\fff{\baselineskip=\normalbaselineskip}
\def\gam{\gamma}
\def\jj{{\bf j}}

\preprint{Imperial/TP/96-97/51}
\title{Generic junction conditions in brane-world scenarios}
\author{Richard A. Battye{$^{1}$} and Brandon Carter{$^{2}$}}
\address{${}^1$ Department of Applied Mathematics and Theoretical Physics,
Centre for Mathematical Sciences, Univeristy of Cambridge,\\
Wilberforce Road, Cambridge CB3 OWA, U.K. \\ ${}^2$ Department d'Astrophysique et de Cosmologie. Centre
National de la Recherche Scientifique, \\ Observatoire de Paris, 92195
Meudon Cedex, France.}
\maketitle
\begin{abstract}
We present the generic junction conditions obeyed by a co-dimension
one brane  in an arbitrary background spacetime. As well as the usual
Darmois-Israel junction conditions which relate the discontinuity in
the extrinsic curvature to the to the energy-momentum tensor of matter
which is localized to the brane, we point out that another condition
must also be obeyed. This condition, which is the analogous to
Newton's second law for a point particle, is trivially satisfied when
$Z_2$ symmetry is enforced by hand, but in more general circumstances
governs the evolution of the brane world-volume. As an illustration of
its effect we compute the force on the brane due to a form field.
\end{abstract}

\date{\today}

\pacs{PACS Numbers : }

\section{Introduction}

Motivated by developments in string theory interest has recently focused on the idea that the 4D universe that we
see is an embedding of a 3-brane in higher dimensional
space-time, the general concept being  known as a brane-world.
In the most popular of these models the universe is
hyper-surface, hyper-brane or co-dimension one object in a 5-dimensional
background~\cite{RS1,RS2}. The ordinary matter is
localized as a distributional source, with an
additional assumption that the extra dimension, which is also non
compact, is $Z_2$ symmetric; both ideas being having phenomenological 
roots in  string
theory scenarios in the more formal contexts of D-branes~\cite{pol} and
orbifolds~\cite{witten} respectively. An enormous amount of effort has
been directed toward understanding how the gravity acts in a
cosmological setting within the framework of these models
using coordinate dependent~\cite{bdl,coorddep} and
coordinate independent~\cite{coordindep} approaches. Central to all
treatments of this problem are the junction conditions which must be
imposed across the discontinuity induced by the brane.

Consider the general case of  a $p$-brane ($p$ is the number of spatial
dimensions of the brane, which has a $(p+1)$-dimensional world-volume)
in an $n$-dimensional spacetime. The generic dynamical
properties of such an object can be treated in a relatively
straightfoward way (see, for example, ref.~\cite{Pey00}) so long as
the evolution is `passive', that is, it takes place
in a background which is fixed and
unaffected by the presence of the brane, as would be the case to a very
good approximation when the relevant coupling (gravitational,
electromagnetic, or other) are sufficiently weak. In contrast if the
`active' effect of such couplings is  sufficiently strong,
there will be  awkward singularity problems involving
divergences which  require non-trivial regularisation
procedures whenever the  co-dimension, $n-p-1$, is  equal to 2 or more,
as in the recently  clarified case~\cite{CB98} of a self-gravitating
string (1-brane) in ordinary 4-dimensional spacetime.

The specific case of co-dimension one ($n=p+2$), the hyper-surface or
hyper-brane which includes the case of a 3-brane in 5 dimensions,
is much simpler, since it has the convenient feature
that its `active' effect on the background does not give rise to
divergences and can be allowed for 
in straightforward way in terms of simple discontinuities in the relevant
field gradients.  Of most interest in the context of
General Relativity is the discontinuity in the space-time metric
$g_{\mu\nu}$ which has been discussed in many works. In this paper
we follow the approach presented in ref.~\cite{MTW} based on a
treatment  pionneered by Darmois~\cite{Darmois27}, and finally cast
into a conveniently coordinate independent form in terms of the
tensorially well defined second fundamental form (or extrinsic curvature
tensor) by Israel~\cite{Israel66} in what is commonly quoted as the
definitive reference on this topic.

The present article has been motivated by the observation made in
ref.~\cite{MTW} that these Darmois-Israel junction conditions are
incomplete since  they only prescribe the active source effect of the
hyper-brane  on the background, and do not include the equation
governing the passive motion of the world-volume within the
background. This passive equation of motion is often derived in
particular coordinate systems for diverse applications (see, for example,
ref.~\cite{guth})  and in the
specific case of $Z_2$ symmetry, often used in brane-world models, it 
is trivial~\cite{bdl}. The main result of this paper is to cast this equation,
which is analogous to Newton's 2nd law, in a coordinate independent
tensorial form in an essentially  equivalent way to that done by
Israel for the case of the active effect. 

Such form for the passive effect has so far been available only for
the weakly coupled limit, for which~\cite{Pey00} it takes the simple
form
\be
\overline T{^{\mu\nu}} K_{\mu\nu}=f\,,\label{weak}
\ee 
where $\overline T{^{\mu\nu}}$ is the energy-momentum density tensor for the
matter and vacuum energy localized on the brane, $K_{\mu\nu}$ is the
second fundamental form or extrinsic curvature tensor, and $\overline
f$ is the magnitude of any orthogonal force density due to
external fields in the direction normal to the brane. Interestingly
the form of (\ref{weak}) is that of Newton's 2nd law, $\overline
T{^{\mu\nu}}$ playing the role of mass and $K_{\mu\nu}$ that of
acceleration, and the more general form which we will derive here will
just replace extrinsic curvature and the extrenal force with their
sectional averages over a small neighbourhood around  the brane.

In fact (\ref{weak}) is a specialization of the more general equation 
$\overline T{^{\mu\nu}}K_{\mu\nu}{^{\rho}}=\perp^{\rho}{_{\nu}}f^{\nu}$,
which applies to weakly coupled branes of any co-dimension~\cite{C95}, where
$K_{\mu\nu}{^{  \rho}}$ is the generalization of the second
fundamental form and the $f^{\nu}$ is the external force
density. Within this formalism $\eta_{\mu\nu}$ is the fundamental form
(or metric) of the brane and its orthogonal complement is defined by 
$\perp_{\mu\nu}=g_{\mu\nu}-\eta_{\mu\nu}$.
If one specializes to the co-dimension one case then the orthogonal projection can be expressed in terms of a single normal vector $n_{\mu}$ in the direction of the extra dimension; the orthogonal complement taking the form 
$\perp_{\mu\nu}=n_\mu n_\nu\,$.
The orthogonal unit vector can also be used to specify the orthogonal
force density magnitude, and the second  fundamental tensor
according to the prescriptions  $\overline f=n_\nu \overline
{f^\nu}$  and $K_{\mu\nu}=K_{\mu\nu}{^\rho}n_\rho$. Equivalently,
without  reference to the 3-index second fundamental tensor, the
extrinsic curvature tensor can be expressed directly 
\be 
K_{\mu\nu}=-\eta^{\rho}{_{\nu}}\overline\nabla_\mu n_\rho\,,
\label{defnext}
\ee
in terms of the tangentially projected differentiation operator,
$\overline\nabla_{\nu}=\eta^\nu{_{\mu}}\nabla_{\nu}$. For the rest of
this article we will work with the 2-index version of the second
fundamental form as is sensible in the case of co-dimension one
objects under consideration here. Nonetheless, we would like to point
out that the notation and some of the methods we use are applicable more generally.

\section{Active gravitational source equation --- the Darmois-Israel formula}

We start by introducing notation that is applicable to a generic field
quantity $Q$ say, before applying it to the specific case of the
extrinsic curvature.  The field is assumed  to vary
smoothly within a neighbourhood $\zeta_{_-} < \zeta < \zeta_{_+}$  of
a timelike hypersurface that is identified  as the
world-volume of the brane under consideration. One can assume that the
brane  is situated at the locus $\zeta=0$ with respect to coordinates
chosen so that $\zeta$  measures the normal distance from the
world-volume, and hence the unit normal will be given by $n_\nu=
\partial_\nu \zeta/(\partial_\mu\zeta\partial^\mu\zeta)^{1/2}$.
We are concerned with limit configurations for
which the  normal derivative $Q^\prime=\partial Q/\partial
\zeta=n^\nu\partial_\nu Q$ becomes extremely large compared with the
corresponding tangential gradient components in the region under
consideration. Therefore, the effective discontinuity as perceived 
on a scale large compared with the  thickness
$\zeta_{_+}-\zeta_{_-}$ will be given by $[Q]=Q^{+}-Q^{-}$.  This
discontinuity will also be expressible as $[Q] =\overline{Q^\prime}$
where  $\overline{\partial_\nu Q} \simeq n_\nu [Q]$, using an
overline for the sectional integral
$\overline{Q^\prime}=\int_{\zeta_-}^{\zeta_+} Q^\prime\ d\zeta$.
Conversely, this is  equal to the coefficient of the Dirac
distribution to which this derivative will tend in the ultra-thin
limit, that is, $Q^\prime$ $=\overline{Q^\prime}\, \delta[\zeta]$. If
we now define $\langle Q\rangle = (Q^{+}+Q^{-})/2$, we see that the
sectional integral of the product $Q Q^\prime$ can be expressed as
$\overline{Q Q^\prime}= \langle Q\rangle [Q]$.

In such a thin brane limit, when the transverse derivatives are
considered as negligibly small compared with normal ones, then
according  to the standard analysis~\cite{MTW} the dominant
contribution to the normal derivative of the second fundamental form
of the constant $\zeta$ hyper-surfaces will be given by
\be
K^\prime_{\mu\nu}\simeq \eta^\kappa{_{\mu}}\eta^\lambda{_{\nu}}
\perp^{\rho}{_{\sigma}}{\cal R}_{\rho\kappa}{^\sigma}{_\lambda}\,,
\label{extprime}
\ee
where ${\cal R}_{\rho\mu}{^\sigma}{_\nu}$ are the
components of the Riemann tensor of the n-dimensional background
geometry and the symbol $\simeq$ denotes equality to the dominant
contribution.
Contraction of this formula gives the relation
$K^\prime\simeq \perp^{\rho}{_{\sigma}}{\cal R}^\sigma{_{\rho}}$,
where $K=K^\nu{_{\nu}}$ is the extrinsic curvature scalar
and ${\cal R}_{\mu\nu}={\cal R}_{\sigma\mu}{^\sigma}{_\nu}$ is the
background Ricci tensor. Taking account of the fact that,
even in the thin limit
for which the background curvature becomes very large, the intrinsic
curvature of the world-volume will remain relatively negligible, it can
be seen that (\ref{extprime}) also implies
\be
K^\prime_{\mu\nu}\simeq \eta^\kappa{_{\mu}}\eta^\lambda{_{\nu}} {\cal
R}_{\kappa\lambda}\,.
\ee
Thus, one can deduce that the dominant components of
the background Ricci tensor will be given  by the asymptotic formula
\be
{\cal R}_{\mu\nu}\simeq K^\prime_{\mu\nu}+ K^\prime\perp_{\mu\nu}\,,
\ee
whose contraction gives the expression ${\cal R}\simeq
2K^\prime$ for the background Ricci scalar ${\cal R}={\cal
R}^\nu{_{\nu}}$.

For physical applications one is usually concerned with the Einstein
tensor ${\cal G}_{\mu\nu}={\cal R}_{\mu\nu}-{1\over 2}g_{\mu\nu}{\cal
R}$ whose dominant components, which are exclusively
tangential, are given by 
\be
{\cal G}_{\mu\nu}\simeq K^\prime_{\mu\nu}-K^\prime\eta_{\mu\nu}\,.
\label{eintensor}
\ee
The active effect of the brane --- the Darmois-Israel formula --- can
then be deduced by performing the sectional integral of this quantity.
Using the notation conventions introduced above, one can deduce that 
\be
\overline{\cal G}_{\mu\nu}= [K_{\mu\nu}]-[K]\eta_{\mu\nu}\,.
\label{israel}
\ee

In $n$-dimensions the Einstein equations take the form
\be
{\cal G}_{\mu\nu}= (n-2) \Omega^{[n-2]}{\rm G}^{[n]}_{\rm N}
T^{\mu\nu} - \Lambda^{[n]}g_{\mu\nu}\,,
\ee
where $\Omega^{[n-2]}$ denotes the surface area of the unit
$(n-2)$-sphere, $G^{[n]}_{\rm N}=M^{2-n}$ is the $n$-dimensional
analogy of the Newton's constant defined in terms of mass scale M,
and $\Lambda^{[n]}$ is the background cosmological constant. In 4-dimensional
spacetime, $\Omega^{[2]}=4\pi$ and $M=M_{\rm pl}$ is the standard
Planck mass, while in the 5-dimensional case of most interest here,
$\Omega^{[3]}=2\pi^2$ and $M$ is the corresponding mass scale, which
might be related to $M_{\rm pl}$ in order to solve the hierarchy problem.
By performing the relevant sectional integration, and substituting in
(\ref{israel}) one can deduce the standard form of the Darmois-Israel
conditions 
\be
[K_{\mu\nu}]={\Omega^{[n-2]}\over M^{n-2}}
\bigg((n-2) \overline T{_{\mu\nu}} -\overline
T^{\rho}{_{\rho}}\eta_{\mu\nu}\bigg)\,,
\ee
for the gravitational effect of the sectionally
integrated energy-momentum density tensor $\overline
T{^{\mu\nu}}=\int_{\zeta_-}^{\zeta_+} T^{\mu\nu}\,d\zeta$\,.  

\section{Passive world-volume evolution equation}

Having re-derived this well-known formula governing the `active'
gravitational source, we now turn our attention to the previously overlooked
question of the corresponding equation governing the dynamical
evolution of the brane world-volume, which can be obtained from the
local dynamical equation
\be
\nabla_{\mu} T^{\mu\nu}=f^\nu\,,
\label{dyneqn}
\ee
where $f^\nu$ is the
force density resulting from coupling to external fields not
taken into account in the evaluation of the  locally
concentrated energy-momentum  density $T^{\mu\nu}$. The sectional
integral of $T^{\mu\nu}$ gives the surface energy-momentum density $\overline
T{^{\mu\nu}}$ which, in the ultra-thin limit, 
is given by $T^{\mu\nu}= \overline
T{^{\mu\nu}}\, \delta[\zeta]$. It follows immediately from (\ref{dyneqn})
that the orthogonal force density $f=n_\nu f^\nu$ will be given
by
\be
f=\nabla_{\mu}\big(T^{\mu\nu}n_\nu\big)-
T^{\mu\nu}\nabla_{\mu}n_\nu \,,
\ee
via a standard manipulation.

Since the Einstein tensor (\ref{eintensor}) has a tangential form and
is proportional to $T^{\mu\nu}$, it can be seen from
(\ref{defnext}) that the last  term can be expressed as 
$T^{\mu\nu}\nabla_{\mu}n_\nu\simeq  -T^{\mu\nu}
K_{\mu\nu}$, while the first term will give a
vanishing surface contribution by Green's theorem 
when we integrate to obtain the total
sectional force density $\overline f=\int_{_-}^{_+}f d\zeta$. This is
given by
\be
\overline f=\overline{T^{\mu\nu}K_{\mu\nu}}\,.
\ee
It can be seen from (\ref{israel})
that the dominant contribution to $T^{\mu\nu}$ is
proportional to $\big(g^{\mu\rho}g^{\nu\sigma} -\eta^{\mu\nu}
g^{\rho\sigma}\big)K^\prime_{\rho\sigma}$, so substituting
$K_{\mu\nu}$ for $Q$ in the general formula   $\overline{Q^\prime
Q}=[Q] \langle Q\rangle$ we finally obtain the required discontinuous
generalisation of (\ref{weak}) in the form
\be
\overline T{^{\mu\nu}}\langle K_{\mu\nu}\rangle=\overline f \,,
\label{evolution}
\ee
where we recall that  $\langle K_{\mu\nu}\rangle
={1\over 2}\big(K^{_+}_{\mu\nu} +K^{_ -}_{\mu\nu})$ denotes the
average of the values of the second fundamental form on the two sides.

To actually apply the evolution equation (\ref{evolution}) it is
necessary to know enough about the intrinsic mechanics of the brane to
be able to evaluate its surface energy-momentum density 
$\overline T{^{\mu\nu}}$ and 
the force term $\overline f$, which will generally be attributable to
the action of  background fields whose energy-momentum density
$T_{_{\rm ba}}^{\,\mu\nu}$ will  provide a bounded contribution to the
total stress energy density distribution $T_{_{\rm to}}^{\,\mu\nu}$
$=T_{_{\rm ba}}^{\,\mu\nu}+\overline T{^{\mu\nu}}\,\delta[\zeta]$.
The requirement that this total should satisfy the divergence
condition $\nabla_{\mu}T_{_{\rm to}}^{\mu\nu}$ $=0$ implies by
(\ref{dyneqn}) that the force density term should be given by $
f=n_\nu f^\nu$ with $f^\nu= -\nabla_{\mu} T_{_{\rm
ba}}^{\,\mu\nu}$.  In the ultra-thin limit this gives $f^\nu\simeq
-n_\mu T_{_{\rm ba}}^{\prime\mu\nu}$ and hence, the
corresponding sectional integral is obtained as $ \overline f{^\nu}
= -n_\mu[T_{_{\rm ba}}^{\,\mu\nu}]$.  Using this to evaluate
the scalar force density $\overline f=n_\nu\overline
f{^\nu}$, we finally obtain the world-volume  evolution equation in the
generically valid form
\be
\overline T{^{\mu\nu}}\langle K_{\mu\nu}\rangle + [T_{_{\rm
ba}}^{\,\mu\nu}]\perp_{\mu\nu}=0\,.
\label{finaleqn}
\ee

\section{Minimal gauge coupling}

We have already commented that $Z_2$ symmetry is often assumed in
brane-world models motivated by the concept of orbifolds, and that the
evolution equation which we have derived is trivially satisfied in
this case. It is, however, interesting to consider what happens if
this assumption is relaxed. A recent generalisation~\cite{DDPV00}
envisages the spontaneous violation of the $Z_2$ reflection symmetry
even without the intervention of an external force. Here,
we consider the obvious further generalisation in which violation
of reflection symmetry is not spontaneous, but imposed as a physical
necessity by the presence of an external force of the kind that will
arise from a minimal gauge field coupling.

The simplest kind of external field coupling that one might imagine is
an antisymmetric $r$-form Ramond gauge field ${A^{\{r\}}}_{\nu_1 ... \nu_r}$. 
For any such field there will be a corresponding gauge independent
field defined as its exterior derivative by
\be
{F^{\{r+1\}}}_{\nu_0\nu_1 ... \nu_{r}}=(r+1)\nabla_{[\nu_0} {A^{\{r\}}}_{\nu_1 ... \nu_{r}]}\,,
\ee
where the square brackets denote antisymmetrization.
The background action associated with such a field, will comprize of
two terms, an external contribution --- usually due to the kinetic energy
term associated with the field --- and coupling term, ${\cal I}_{_{\rm
ba}}={\cal I}_{_{\rm ex}} + {\cal I}_{_{\rm co}}$. One can write the
external part in terms of an $n$-dimensional integral of a Lagrangian
density ${\cal I}_{_{\rm ex}}=\int{\cal
L}_{_{\rm ex}}\,\Vert g\Vert^{1/2}d^nx$ with the standard Lagrangian density taking the quadratic form
\be
{\cal L}_{_{\rm ex}}= -{m^{n-2-2r}\over
2(r+1)! \Omega^{[n-2]}}F^{\{r+1\}\,\nu_0 ...\nu_r}
{F^{\{r+1\}}}_{\nu_0 ... \nu_r}\,,
\ee
where
$m$ is a fixed coupling constant having the
dimensions of mass, whose presence can be seen to be redundant in the
most familiar example, namely the case of Maxwellian electrodynamics
characterised by r=1 with n=4, but not in more general situations.
The corresponding external energy-momentum tensor ${{T_{_{\rm ex}}}^{\mu}}_{\rho}$ will be given by
\be
{{T_{_{\rm ex}}}^{\mu}}_{\rho}=2g_{\nu\rho}{\partial
{\cal L}_{_{\rm ex}}\over\partial g_{\mu\nu}}+{g^{\mu}}_{\rho}{\cal L}_{_{\rm
ex}}={m^{n-2-2r}\over  r! \Omega^{[n-2]}}\Big(F^{\{
r+1\}\,\mu\nu_1 ...\nu_r} {F^{\{r+1\}}}_{\rho\nu_1
...\nu_r} -{1\over 2({r+1})}\,{g^\mu}_{\rho}F^{\{r+1\}\,\nu_0
...\nu_r} {F^{\{r+1\}}}_{\nu_0 ... \nu_r}\Big)\,.
\label{emtensor}
\ee

The standard coupling of such a field to a $p$-brane is using an
action ${\cal I}_{_{\rm co}}$ of the generalized Wess-Zumuno type 
\be
{\cal I}_{_{\rm co}}=\int\overline{{\cal L}_{_{\rm
co}}}\,
\Vert\gamma\Vert^{1/2}\ d^{p+1} \sigma\,,
\label{coupleaction}
\ee
which will require that $r=p+1$. The Lagrangian surface density has the
form
\be
\overline{{\cal L}_{_{\rm co}}} ={1\over (p+1)!}\,
\overline{\jj}^{\{p+1\}\,\nu_1 ... \nu_{p+1}} {A^{\{p+1\}}}_{\nu_1
... \nu_{p+1}}\,,
\label{couplelag}
\ee
with
\be
{\overline{\jj}}^{\{p+1\}\,\nu_1
... \nu_{p+1}} = {e^{\{p+1\}}}{\cal E}^{\nu_1 ... \nu_{p+1}}\,,
\label{minimal}
\ee
where ${\cal E}^{\nu_1 ... \nu_{p+1}}$, normalized such that 
${\cal E}^{\nu_1 ...\nu_{p+1}}{\cal E}_{\nu_1
...\nu_{p+1}} = -(p+1)!$, is the usual antisymmetric surface
element, $\Vert\gamma\Vert$ is the modulus of the determinant of
the induced metric given with respect to internal coordinates
$\sigma^a$  ($a$=1, ... , p+1) by $\gamma_{ab}=$
$\eta_{\mu\nu}\partial_ax^\mu
\partial_bx^\nu=g_{\mu\nu}\partial_ax^\mu\partial_bx^\nu$,
and $e^{\{p+1\}}$ is a dimensionless charge coupling constant.

One can formally convert the world-volume integral (\ref{coupleaction}) to the
background spacetime integral
\be
{\cal I}_{_{\rm co}}=\int \jj^{\{p+1\}\,\nu_1
... \nu_{p+1}} {A^{\{p+1\}}}_{\nu_1 ... \nu_{p+1}}\,\Vert g\Vert^{1/2}d^nx\,,
\ee
by the introduction of a Dirac distribution with respect to the
world-volume locus $x^\mu= \bar x{^\mu}\{\sigma\}$, and hence the
source flux vector is given by 
\be
\jj^{\{p+1\}\,\nu_1 ... \nu_{p+1}} =\Vert
g\Vert^{-1/2}\int\delta^{(n)}[x-\bar
x\{\sigma\}] \,
{\overline\jj}^{\{p+1\}\,\nu_1 ... \nu_{p+1}}\,
\Vert\gamma\Vert^{1/2} d^{p+1}\sigma\,.
\label{denssurf}
\ee
The field equation can then be derived directly from this,
\be
m^{n-2p-4}\nabla^{\mu}{F^{\{p+2\}}}_{
\mu\nu_1 ... \nu_{p+1}} = - \Omega^{[\rm n-2]}\,{\jj^{\{p+1\}}}_{
\nu_1 ... \nu_{p+1}}\,,
\label{fieldequation}
\ee
and (as shown elsewhere~\cite{C95}) the corresponding generalised
Faraday-Lorentz force density law is given by
\be
f_\mu={1\over(p+1)!}\,{F^{\{p+2\}}}_{\mu\nu_1 ... \nu_{p+1}}\,
{\jj^{\{p+1\}\nu_1 ... \nu_{p+1}}}\,.
\label{lorentz}
\ee
These expressions are valid for a general $p$-brane in an
$n$-dimensional spacetime. For a co-dimension larger than one such a
source gives rise to a divergence, for example the n=4, d=2 axion
divergence in strings (see, for example, ref~\cite{BS}), although this may be cancelled by the inclusion of an
appropriately weighted dilaton field~\cite{Pey98}. 

This kind of divergence problem does not arise in the co-dimension one
case with which we are concerned here. In this case 
resulting field ${F^{\{n\}}}_{\nu_0 ... \nu_{n-1}}$ will
remain bounded, with just a discontinuity given according  to
(\ref{fieldequation}), 
by the limit of the asymptotic formula
$m^{-n} {F^{\{n\}\prime}}_{\
\mu\nu_1 ... \nu_{n-1}}\simeq - n\Omega^{[\rm n-2]}
n_{[\mu}\,{\jj^{\{n-1\}}}_{\nu_1 ... \nu_{n-1}]}$. Computing the
sectional integeral, gives the following formula for the discontinuity
\be
m^{-n} [{F^{\{n\}}}_{\mu\nu_1
... \nu_{n-1}}] = - n\Omega^{[\rm n-2]}n_{[\mu}\,
{\overline{\jj}^{\{n-1\}}}_{\nu_1 ... \nu_{n-1}]}\,,
\ee
with ${\perp^{\mu}}_{\nu_1}\,
\overline{\jj}^{\{n-1\}\,\nu_1 ... \nu_{n-1}}=0$. In the ultra-thin limit
(\ref{denssurf}) can replaced by the simpler formula 
\be
\jj^{\{n-1\}\,\nu_1 ... \nu_{n-1}}
={\overline{\jj}}^{\{n-1\}\,\nu_1 ... \nu_{n-1}}\,\delta[\zeta]\,,
\ee
and it can be seen from (\ref{lorentz}) that the
surface force density will be given in terms of the mean field
$\langle {F^{\{n\}}}_{\nu_0 ... \nu_{n-1}}\rangle$ by
\be
\overline f_\mu={1\over (n-1)!}\,\langle {F^{\{n\}}}_{
\mu\nu_1 ... \nu_{n-1}}\rangle\,{\overline
{\jj}}^{\{n-1\}\,\nu_1 ... \nu_{n-1}}\,.
\ee

Since there exists just one totally antisymmetric $n$-vector in an
$n$-dimensional spacetime, one can write the  corresponding physical
field ${F^{\{n\}}}_{\nu_0 ... \nu_{n-1}}$ in
terms of a pseudo-scalar magnitude $F^{\{n\}}$ given by
\be
{F^{\{n\}}}_{\nu_0 ... \nu_{n-1}}= F^{\{n\}}\,\epsilon_{\nu_0
... \nu_{n-1}}\,,
\ee
where
$\epsilon_{\nu_0 ... \nu_{n-1}}$ is the background spacetime
antisymmetric $n$-form normalized in an equivalent way to ${\cal
E}^{\nu_1 .. \nu_{p+1}}$. Therefore, the external energy-momentum tensor
(\ref{emtensor}) will reduce to the form
\be
{{T_{_{\rm ex}}}^{\mu}}_{\rho}= - {1\over  2 m^n\Omega^{[n-2]}}
\big(F^{_{\{\rm n\}}}\big)^2 {g^\mu}_{\rho}\,,
\label{finalstress}
\ee
since $\epsilon^{\mu\nu_1 ... \nu_{n-1}}{\cal E}_{\rho\nu_1
... \nu_{n-1}}=-(n-1)!\,{g^{\mu}}_{\rho}$ and the field equation (\ref{fieldequation}) will reduce to the form
\be
\nabla_{\mu} F^{_{\{n\}}} = {m^n\Omega^{[n-2]}\over (n-1)!}
\epsilon_{\mu\nu_1 ... \nu_{\rm n-1}}\,
\jj^{\{n-1\}\,\nu_1 ... \nu_{n-1}}\,,
\ee
which implies that in a source free vacuum the
value of the pseudo scalar $F^{\{n\}}$ will be constant,so that
the effect of (\ref{finalstress}) will be
perceived just as a positive increment to the cosmological
constant. This increment will, however, be subject to a discontinuous
change across the brane world-volume, where it can be seen from the
minimal coupling condition (\ref{minimal}) that we shall have
\be
[F^{\{n\}}]= e^{\{n-1\}} m^n\Omega^{[\rm n-2]}\,,
\ee 
subject to the understanding that the
orientation of the brane is such that the unit normal is given by $n_\mu=\epsilon_{\mu\nu_1 ... \nu_{n-1}} {\cal
E}^{\nu_1 ...\nu_{n-1}}/(n-1)!$.

The product $\Vert g\Vert^{1/2}{\cal L}_{_{\rm co}}$ is independent of
the metric and so gives no contribution to the energy-momentum
density, $T_{_{\rm co}}^{\,\mu\nu}=0$ and hence that the
background  contribution in the world-volume evolution equation
(\ref{finaleqn}) will  be given simply by  ${T_{_{\rm ba}}}^{\mu\nu}$
$={T_{_{\rm ex}}}^{\mu\nu}$. Therefore, 
the required force density $\overline f$
$=-n_\mu n_\nu [T_{_{\rm ba}}^{\,\mu\nu}]$ can thus be seen
from (\ref{finalstress}) to be given by the constant value
\be
\overline f=  {1\over  2 m^n \Omega^{[\rm
n-2]}}\big[\big(F^{_{\{ n\}}}\big)^2\big] = e^{\{n-1\}}
\langle F^{_{\{\rm n\}}}\rangle\,.
\ee

\section{Summary}

We have derived in a coordinate independent way the passive effect of
the world-cvolume evolution on a co-dimension one brane. This equation
is interesting in the context of brane-world models when $Z_2$
symmetry is not imposed by hand and we have illustrated how such a
possibility might manifest itself when the brane is minimally coupled
to an $(n-1)$-form field. We have showed that such a field is
constant except for a discontinuity at the
brane and, hence, 
the effective background cosmological constant will be different
on either side of the brane giving us a realistic situation in which
the models discussed in ref.\cite{DD} are valid. The formulae
derived in this paper can be used to compute, for example, the
evolution of the scale-factor in a FRW type model for the universe
confined to a 3-brane in 5 spacetime dimensions~\cite{CU}. 

\renewcommand{\thefootnote}{\arabic{footnote}} \setcounter{footnote}{0}

\def\jnl#1#2#3#4#5#6{\hang{#1, {\it #4\/} {\bf #5}, #6 (#2).} }
\def\jnltwo#1#2#3#4#5#6#7#8{\hang{#1, {\it #4\/} {\bf #5}, #6; {\it
ibid} {\bf #7} #8 (#2).}}  \def\prep#1#2#3#4{\hang{#1, #4.} }
\def\proc#1#2#3#4#5#6{{#1, in {\it #4\/}, #5, eds.\ (#6,#2).} }
\def\book#1#2#3#4{\hang{#1, {\it #3\/} (#4, #2).} }
\def\jnlerr#1#2#3#4#5#6#7#8{\hang{#1, {\it #4\/} {\bf #5}, #6 (#2).
{Erratum:} {\it #4\/} {\bf #7}, #8.} }  \def\prl{Phys.\ Rev.\ Lett.}
\def\pr{Phys.\ Rev.}  \def\pl{Phys.\ Lett.}  \def\np{Nucl.\ Phys.}
\def\prp{Phys.\ Rep.}  \def\rmp{Rev.\ Mod.\ Phys.}  \def\cmp{Comm.\
Math.\ Phys.}  \def\mpl{Mod.\ Phys.\ Lett.}  \def\apj{Ap.\ J.}
\def\apjl{Ap.\ J.\ Lett.}  \def\aap{Astron.\ Ap.}  \def\cqg{Class.\
Quant.\ Grav.}  \def\grg{Gen.\ Rel.\ Grav.}  \def\mn{MNRAS}
\def\ptp{Prog.\ Theor.\ Phys.}  \def\jetp{Sov.\ Phys.\ JETP}
\def\jetpl{JETP Lett.}  \def\jmp{J.\ Math.\ Phys.}  \def\zpc{Z.\
Phys.\ C} \def\cupress{Cambridge University Press} \def\pup{Princeton
University Press} \def\wss{World Scientific, Singapore}
\def\oup{Oxford University Press}

\pagebreak
\pagestyle{empty}


\begin{thebibliography}{99}

\bibitem{RS1}
\jnl{L. Randall and R. Sundrum}{1999}{}{\prl}{83}{3370}

\bibitem{RS2}
\jnl{L. Randall and R. Sundrum}{1999}{}{\prl}{83}{4690}

\bibitem{pol}
\jnl{J. Polchinski}{1995}{}{\prl}{75}{4724}

\bibitem{witten}
\jnl{E. Witten}{1996}{}{\np}{B471}{135}

\bibitem{bdl} 
\jnl{P. Bin\'etruy, C. Deffayet,
D. Langlois}{2000}{Non conventional cosmology from a brane
universe}{\np}{B565}{269}

\bibitem{coorddep}
\jnl{J.M. Cline, C. Grojean and G. Servant}{1999}{}{\prl}{83}{4245}
\jnl{C. Csaki, M. Graesser, C. Kolda and
J. Terning}{1999}{}{\pl}{B462}{34}
\jnl{P. Bin\'etruy,
C. Deffayet, U. Ellwanger and
D. Langlois}{1999}{}{\pl}{B477}{285}\prep{E.E. Flannagan, S.-H. Tye and I. Wasserman}{1999}{}{hep-ph/9910498}

\bibitem{coordindep}
\jnl{T. Shiromizu, K. Maeda and
M. Sasaki}{2000}{}{\pr}{D62}{045915}\prep{K. Maeda and
D. Wands}{2000}{}{hep-th/0008188}\prep{A. Mennim and R.A. Battye}{2000}{}{hep-th/0008192}

\bibitem{Pey00}
\prep{B. Carter}{2000}{Essentials of classical brane dynamics}{in
proceedings of Peyresq meeting, June 2000, gr-qc/0012036}

\bibitem{CB98} 
\jnl{R.A. Battye and B. Carter}{1995}{}{\pl}{357B}{29}
\jnl{B. Carter, R.A. Battye}{1998}{Non-divergence of gravitational
self interactions for Nambu-Goto strings}{\pl}{B430}{49}

\bibitem{MTW} 
\book{C.W. Misner, K.S. Thorne and  J.A. Wheeler}{1973}{Gravitation}
{Freeman, San Francisco}

\bibitem{Darmois27}
\jnl{G. Darmois}{1927}{Les \'equations de la gravitation einsteinienne}
{\it M\'emorial des Sciences Math\'ematiques}{XXV}
{Gauthier - Villars, Paris}

\bibitem{Israel66} 
\jnlerr{W. Israel}{1966}{Singular hypersurfaces and thin shells in general relativity}{Nuovo Cimento}{B44}{1}{B48}{463}

\bibitem{guth}
\jnl{V.A. Berezin, V.A. Kuzmin and
I.I. Tkachev}{1983}{}{\pl}{120B}{91}\jnl{S.K. Blau, E.I. Guendelman
and A.H. Guth}{1987}{}{\pr}{D35}{1747}

\bibitem{C95} 
\proc{B. Carter}{1995}{Dynamics of cosmic strings and other brane models}
{Formation and interactions of topological defects,
NATO ASI {\bf B349}, Newton Inst., Cambridge 1994}{R. Brandenberger, A.-C. Davis}{Plenum, New York}

\bibitem{DDPV00} 
\prep{A.-C. Davis, S. Davis, W.B. Perkins and  I.R. Vernon}{2000}{Brane world phenomenology and the $Z_2$ symmetry}{hep-ph/0008132}

\bibitem{BS}
\jnl{R.A. Battye and E.P.S. Shellard}{1995}{}{\prl}{75}{4354}
\jnl{R.A. Battye and E.P.S. Shellard}{1996}{}{\pr}{53}{1811}

\bibitem{Pey98} 
\prep{B. Carter}{2000}{Cancellation of linearised axion - dilaton self
- interaction in strings}{in proceedings of Peyresq meeting, June
1998, to be published in {\it Int J. Theor. Phys.}, hep-th/0001136}

\bibitem{DD}
\jnl{N. Deruelle and
T. Dolezel}{2000}{}{\pr}{D62}{10350}\jnl{H. Stoica, S.-H. Tye and I. Wasserman}{2000}{}{\pl}{B482}{205}

\bibitem{CU}
\prep{B. Carter and J.-P. Uzan}{2001}{}{gr-qc/0101010}

\end{thebibliography}
\end{document}